\documentclass[prx,twocolumn,floatfix,superscriptaddress,longbibliography]{revtex4-1}
\usepackage{tikz}
\usepackage{mathtools}
\usetikzlibrary{shapes,arrows,positioning}
\usepackage{standalone}
\usepackage{amssymb,amsmath,amstext}
\usepackage{graphicx}
\usepackage{epstopdf}
\usepackage{color}
\usepackage{appendix}
\usepackage[T1]{fontenc}
\usepackage{bbold}
\usepackage{bbm}
\usepackage{float}
\usepackage{latexsym}
\usepackage{xr-hyper}
\usepackage[colorlinks=true,citecolor=blue,linkcolor=magenta]{hyperref}
\usepackage[latin1]{inputenc}

\usepackage{tikz}
\usetikzlibrary{shapes,arrows}
\renewcommand{\vec}[1]{\boldsymbol{#1}}  
\newcommand{\<}{\langle}
\renewcommand{\>}{\rangle}

\long\def\ca#1\cb{} 

\newcommand{\ket}[1]{|#1\rangle}               
\newcommand{\bra}[1]{\langle #1|}              
\newcommand{\dya}[1]{\ket{#1}\!\bra{#1}}
\newcommand{\dyad}[2]{\ket{#1}\!\bra{#2}}        

\newcommand{\CC}{\mathcal{C}}
\newcommand{\DC}{\mathcal{D}}

\newcommand{\YC}{\mathcal{Y}}
\newcommand{\ZC}{\mathcal{Z}}

\newcommand{\HS}{\text{HS}}

\newcommand{\Tr}{{\rm Tr}}

\renewcommand{\leq}{\leqslant}

\renewcommand{\Re}{\text{Re}}

\renewcommand{\vec}[1]{\boldsymbol{#1}}  
\newcommand{\ot}{\otimes}
\newcommand{\ad}{^\dagger}

\begin{document}

\title{Variational Consistent Histories as a Hybrid Algorithm for Quantum Foundations}

\author{Andrew Arrasmith}
\affiliation{Theoretical Division, MS 213, Los Alamos National Laboratory, Los Alamos, NM 87545, USA.}
\affiliation{Department of Physics, University of California Davis, Davis, CA 95616, USA.}

\author{Lukasz Cincio} 
\affiliation{Theoretical Division, MS 213, Los Alamos National Laboratory, Los Alamos, NM 87545, USA.}

\author{Andrew T. Sornborger} 
\affiliation{Information Sciences, MS 256, Los Alamos National Laboratory, Los Alamos, NM 87545, USA.}

\author{Wojciech H. Zurek} 
\affiliation{Theoretical Division, MS 213, Los Alamos National Laboratory, Los Alamos, NM 87545, USA.}

\author{Patrick J. Coles} 
\affiliation{Theoretical Division, MS 213, Los Alamos National Laboratory, Los Alamos, NM 87545, USA.}

\begin{abstract}
While quantum computers are predicted to have many commercial applications, less attention has been given to their potential for resolving foundational issues in quantum mechanics. Here we focus on quantum computers' utility for the Consistent Histories formalism, which has previously been employed to study quantum cosmology, quantum paradoxes, and the quantum-to-classical transition. We present a variational hybrid quantum-classical algorithm for finding consistent histories, which should revitalize interest in this formalism by allowing classically impossible calculations to be performed. In our algorithm, the quantum computer evaluates the decoherence functional (with exponential speedup in both the number of qubits and the number of times in the history), and a classical optimizer adjusts the history parameters to improve consistency.  We implement our algorithm on a cloud quantum computer to find consistent histories for a spin in a magnetic field, and on a simulator to observe the emergence of classicality for a chiral molecule.
\end{abstract}
\maketitle

\section*{Introduction}
The foundations of quantum mechanics (QM) have been debated for the past century~\cite{Wheeler-Zurek,Foundations2000}, including topics such as the EPR paradox, hidden-variable theories, Bell's Theorem, Born's rule, and the role of measurements in QM. This also includes the quantum-to-classical transition, i.e., the emergence of classical behavior (objectivity, irreversibility, lack of interference, etc.) from quantum laws~\cite{Joos,ZurekReview,Schlosshauer}.

The Consistent Histories (CH) formalism was introduced by Griffiths, Omn\`es, Gell-Mann, and Hartle to address some (though not all) of the aforementioned issues~\cite{Griffiths-Original,Omnes, GH}. One inventor considered CH to be ``the Copenhagen interpretation done right''~\cite{Griffiths-Original}, as it resolves some of the paradoxes of quantum mechanics by enforcing strict rules for logical reasoning with quantum systems. In this formalism, the Copenhagen interpretation's focus on measurements as the origin of probabilities is replaced by probabilities for sequences of events (histories) to occur, and hence by avoiding measurements it avoids the measurement problem. The sets of histories whose probabilities are additive (as the histories do not interfere with each other) are considered to be consistent and are thus the only ones able to be reasoned about in terms of classical probability and logic~\cite{Omnes}.

Regardless of one's opinion of the philosophical interpretation (on which this paper is agnostic), this computational framework has proven useful in applications such as attempting to solve the cosmological measure problem~\cite{Hartle_measure,Lloyd_measure}, understanding quantum jumps~\cite{quantum_jumps}, and evaluating the arrival time for particles at a detector~\cite{Harrival_time,Harrival_time2,arrival_time}. One of the main reasons that this framework has not received more attention and use is that carrying out the calculations for non-trivial cases (e.g., discrete systems of appreciable size or continuous systems that do not admit approximate descriptions by exactly solvable path integrals) can be difficult~\cite{quantum_jumps,Brun_brownian_motion}. While numerical approaches have been attempted~\cite{POHLE1995435,numericalCHandM}, they require exponentially scaling resources as either the number of times considered or the system size grows. This makes classical numerical approaches unusable for any but the simplest cases.

With the impending arrival of the first noisy intermediate-scale quantum computers~\cite{Preskill}, the field of variational hybrid quantum-classical algorithms (VHQCAs), which make the most of short quantum circuits combined with classical optimizers, has been taking off. VHQCAs have now been demonstrated for a myriad of tasks ranging from factoring to finding ground states, among others~\cite{VQE,VQF, farhi2014QAOA, romero2017quantum, li2017efficient, johnson2017qvector, Khatri_LaRose_Poremba_Cincio_Sornborger_Coles_2018, LaRose}. The VHQCA framework potentially brings the practical applications of quantum computers years closer to fruition.

Here we present a scalable VHQCA for the CH formalism. Our algorithm achieves an exponential speedup over classical methods both in terms of the system size and the number of times considered. It will allow exploration beyond toy models, such as the quantum-to-classical transition in mesoscopic quantum systems. We implement this algorithm on IBM's superconducting qubit quantum processor and obtain results in good agreement with theoretical expectations, suggesting that useful implementations of our algorithm may be feasible on near-term quantum devices.

\section*{Results}
\subsection*{Consistent Histories Background}

In the CH framework~\cite{griffiths-book,H_review,newerCHreview}, a history $\YC^{\vec{\alpha}}$ is a sequence of properties (i.e., projectors onto the appropriate subspaces) at a succession of times $t_1 < t_2 < \ldots <t_k$,
\begin{equation}
    \YC^{\vec{\alpha}}=(P_1^{\alpha_1},P_2^{\alpha_2},\ldots, P_k^{\alpha_k})\,,
\end{equation}
where $P_j^{\alpha_j}$ is chosen from a set $P_j$ of projectors that sum to the identity at time $t_j$.
For example, for a photon passing through a sequence of diffraction gratings and then striking a screen, a history could be the photon passed through one slit in the first grating, another slit in the second, and so on. Clearly, we find interference between such histories unless there is some sense in which the photon's path has been recorded. Since there is interference, we cannot add the probabilities of the different histories classically and expect to correctly predict where the photon strikes the screen.

The CH framework provides tools for determining when a family (i.e., a set that sums to the multi-time identity operator) of histories $\mathcal F = \{\YC^{\vec{\alpha}}\}$ exhibits interference, which is not always obvious. In this framework, one defines the so-called class operator
\begin{equation}
    \CC^{\vec{\alpha}}= P_k^{\alpha_k}(t_k)P_{k-1}^{\alpha_{k-1}}(t_{k-1})\ldots P_1^{\alpha_1}(t_1),
\end{equation}
which is the time-ordered product of the projection operators (now in the Heisenberg picture and hence explicitly time dependent) in history $\YC^{\vec{\alpha}}$. If the system is initially described by a density matrix $\rho$, the degree of interference or overlap between histories $\YC^{\vec{\alpha}}$ and $\YC^{\vec{\alpha}'}$ is
\begin{equation}
\label{eq:GGH consistency func}
    \mathcal{D}(\vec{\alpha},\vec{\alpha}')=\Tr\left(\mathcal{C}^{\vec{\alpha}}\rho\, \mathcal{C}^{\vec{\alpha}'\dagger}\right).
\end{equation}
This quantity is called the decoherence functional. The consistency condition for a family of histories $\mathcal F$ is then
\begin{equation}
\label{eq:Real GGH consistency condition}
   \Re( \DC(\vec{\alpha},\vec{\alpha}'))=0 \,,\quad \forall \vec{\alpha}\ne\vec{\alpha}'\,.
\end{equation}
If and only if this condition holds do we say that  $\DC(\vec{\alpha},\vec{\alpha})$ is the probability for history $\YC^{\vec{\alpha}}$. For computational convenience, we will instead work with a stronger condition~\cite{H_review}:
\begin{equation}
\label{eq:GGH consistency condition}
   \DC(\vec{\alpha},\vec{\alpha}')=0 \,,\quad \forall \vec{\alpha}\ne\vec{\alpha}'\,,
\end{equation} 
Since we are presenting a numerical algorithm, it will also be useful to consider approximate consistency, where we merely insist that the interference is small in the following sense:
\begin{equation}
\label{eq:GGH approx consistency condition}
    |\DC(\vec{\alpha},\vec{\alpha}')|^2\leq \epsilon^2 \DC(\vec{\alpha},\vec{\alpha})\DC(\vec{\alpha}',\vec{\alpha}') \,,\quad \forall \vec{\alpha}\ne\vec{\alpha}'\,,
\end{equation}
which guarantees that probability sum rules for $\mathcal F$ are satisfied within an error of $\epsilon$~\cite{approximate_consistency}.

To study consistency arising purely from decoherence (i.e., records in the environment), researchers have proposed a functional that instead takes a partial trace over $\rm{E}$, which is (a subsystem of) the environment~\cite{RZZ,Finkelstein_pt}:
\begin{equation}
\label{eq:FRZZ consistency func}
    \DC_{\rm{pt}}(\vec{\alpha},\vec{\alpha}')=\Tr_{\rm{E}}\left(\mathcal{C}^{\vec{\alpha}}\rho\, \mathcal{C}^{\vec{\alpha}'\dagger}\right).
\end{equation}
With this modification, the consistency condition is
\begin{equation}
    \DC_{\rm{pt}}(\vec{\alpha},\vec{\alpha}')=\vec{0}\,,\quad \forall \vec{\alpha}\ne\vec{\alpha}'\,,
\end{equation}
where $\vec{0}$ is the zero matrix. Instead of only signifying the lack of interference, partial-trace consistency singles out whether or not the records of the histories in the environment interfere. Note that the full-trace condition of Eq.~\eqref{eq:GGH consistency condition} is satisfied when this partial-trace consistency is satisfied, but the converse does not hold~\cite{RZZ}. 
\begin{figure}
    \centering
    \includegraphics[width=\columnwidth]{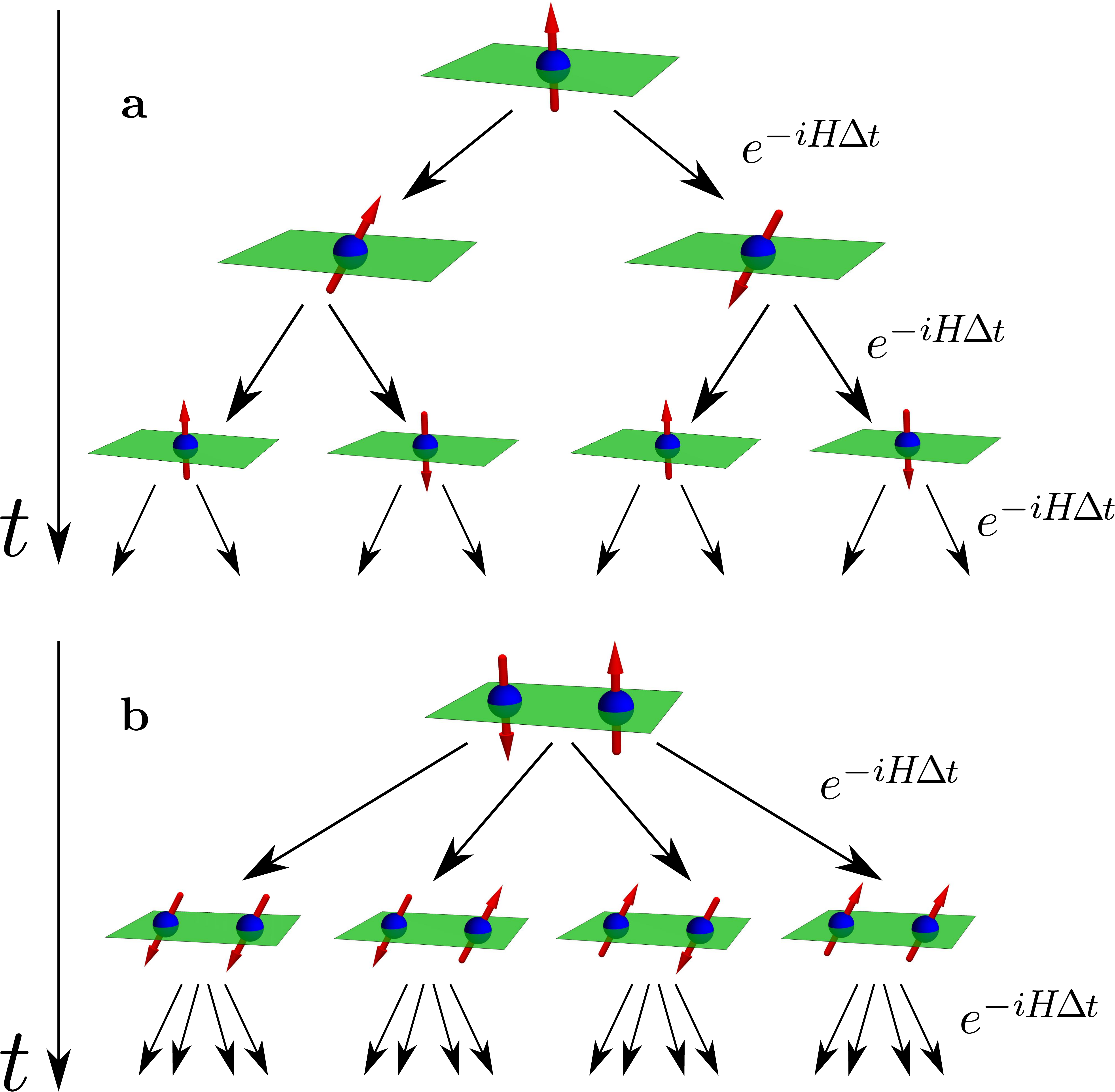}
    \caption{An illustration of the branching of histories for $k$ time steps. A one-spin ($n=1$) and two-spin ($n=2$) system, respectively shown in panels \textbf{a} and \textbf{b}, have $2^k$ and $2^{2k}$ different histories. Here, $k=3$ in \textbf{a} and $k=2$ in \textbf{b}.}
    \label{fig:cartoon}
\end{figure}

With this formalism in hand, we can now see why classical numerical schemes for CH have faced difficulty. For example, consider histories of a collection of $n$ spin-1/2 particles for $k$ time steps, depicted in Fig.~\ref{fig:cartoon}. The number of histories is $2^{nk}$, and hence there are $\sim 2^{2nk}$ decoherence functional elements. Furthermore, evaluating each decoherence functional element $\DC(\vec{\alpha},\vec{\alpha}')$ requires the equivalent of a Hamiltonian simulation of the system, i.e., the multiplication of $2^n \times 2^n$ matrices. This means modern clusters would take centuries to evaluate the consistency of a family of histories with $k=2$ time steps and $n=10$ spins. Given this limitation, we can see why, for the most part, only toy models have been analyzed in this framework thus far. 

\subsection*{Hybrid algorithm for finding consistent histories} \label{sec:VCH}

\begin{figure}[h!t]
    \includegraphics[width=\columnwidth]{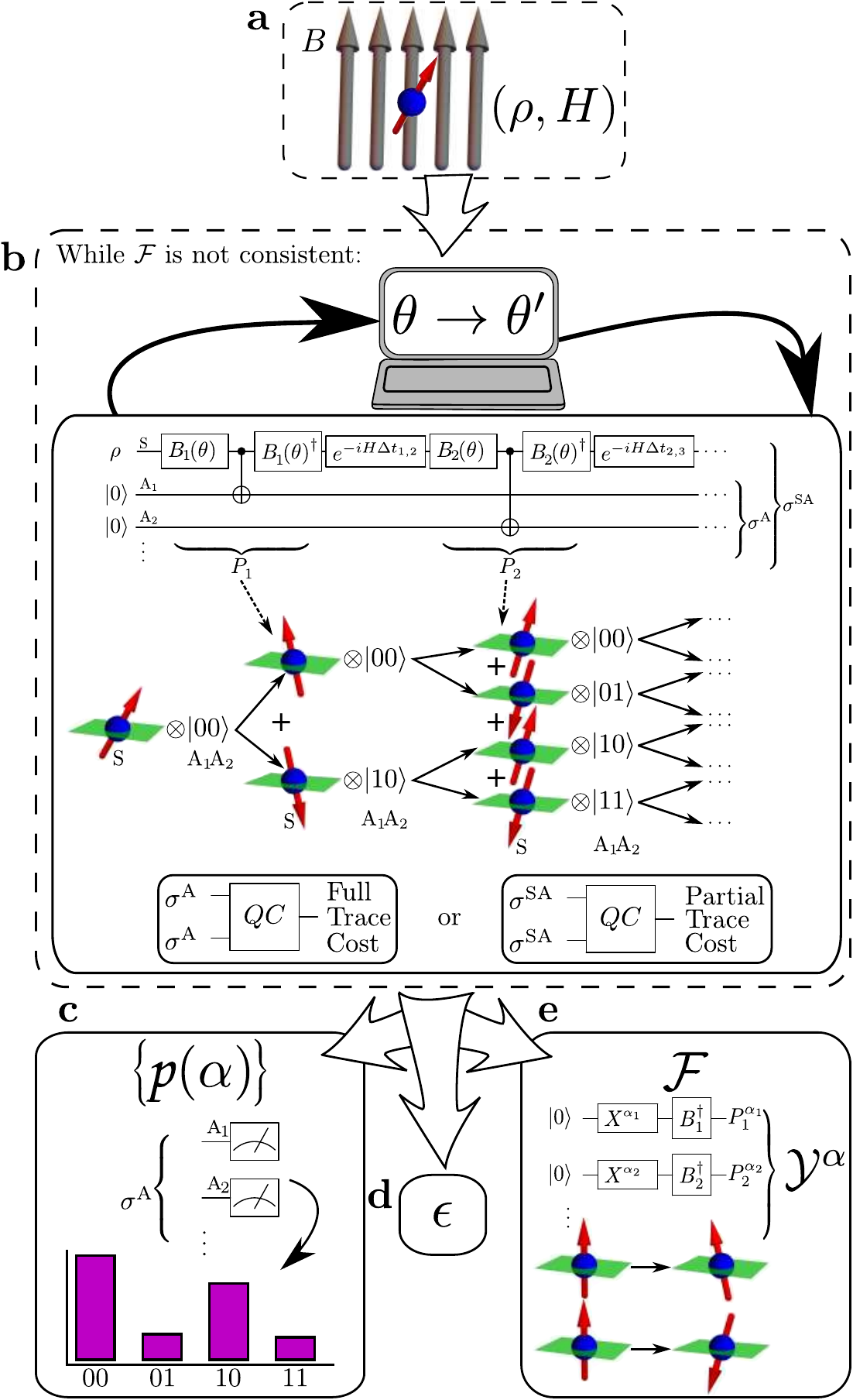}
    \caption{Flowchart for VCH. The goal of VCH is to take a physical model (panel \textbf{a}) and output an approximately consistent family $\mathcal F$ of histories (\textbf{e}), their associated probabilities $\{p({\vec{\alpha}})\}$ (\textbf{c}), and a measure $\epsilon$ of how consistent $\mathcal F$ is (\textbf{d}). This is accomplished via a parameter optimization loop (\textbf{b}), which is a hybrid quantum-classical computation. Here the classical computer adjusts the projector parameters (contained in the gates $\{B_j(\theta)\}$, where $B_j(\theta)$ diagonalizes the $P_j$ projectors) and a quantum computer returns the cost. Note that $P_j$ denotes the set of Schrodinger-picture projectors at the $j^{th}$ time. The optimal parameters are then used to compute the probabilities of the most likely histories in $\mathcal F$ (panel \textbf{c}) and to prepare the projectors for any history in $\mathcal F$ (panel (\textbf{e}), where $X$ is the Pauli-$X$ operator). While the quantum circuits are depicted for a one-qubit system, Appendix~\ref{sct:appA} discusses the generalizations to multi-qubit systems, non-trivial environment $\rm{E}$, coarse-grained histories, and branch-dependent histories.}
    \label{fig:flowchart}
\end{figure}

We refer to our VHQCA as Variational Consistent Histories (VCH), see Fig.~\ref{fig:flowchart}. VCH takes as its inputs a physical model (i.e., an initial state $\rho$ and a Hamiltonian $H$) and some ansatz for the types of projectors to consider. It outputs: (1) a family $\mathcal F$ of histories that is (approximately) full and/or partial trace consistent in the form of projection operators prepared on a quantum computer, (2) the probabilities of the most likely histories $\YC^{\vec{\alpha}}$ in $\mathcal F$, and (3) a bound on the consistency parameter $\epsilon$. 

VCH involves a parameter optimization loop, where a quantum computer evaluates a cost function that quantifies the family's inconsistency, while a classical optimizer adjusts the family (i.e., varies the projector parameters) to reduce the cost. Classical optimizers for VHQCAs are actively being investigated~\cite{mcclean2016theory, LaRose}, and one is free to choose the classical optimizer on an empirical basis.

To compute the cost, note that the elements of the decoherence functional form a positive semi-definite matrix with trace one. In VCH, we exploit this property to encode $\DC$ in a quantum state $\sigma^{\rm{A}}$, whose matrix elements are $\<\vec{\alpha}|\sigma^{\rm{A}}|\vec{\alpha}'\>=\DC (\vec{\alpha},\vec{\alpha}')$. Step \textbf{b} of Fig.~\ref{fig:flowchart} shows a quantum circuit that prepares $\sigma^{\rm{A}}$. (See Appendix~\ref{sct:appB} for more details.) This circuit transforms an initial state $\rho \ot \dya{\vec{0}}$ on systems $\rm{SA}$, where $\rm{S}$ simulates the physical system of interest and $\rm{A}$ is an ancilla system, into a state $\sigma^{\rm{SA}}$ whose marginal is $\sigma^{\rm{A}}$. For the full trace consistency, we introduce a global measure of the (in)consistency that quantifies how far $\sigma^{\rm{A}}$ is from being diagonal, which serves as our cost function:
\begin{equation}
\label{eq:cost}
    C :=\sum_{\vec{\alpha}\ne\vec{\alpha'}} |\DC(\vec{\alpha},\vec{\alpha}')|^2= D_{\HS}(\sigma^{\rm{A}}, \ZC^{\rm{A}}(\sigma^{\rm{A}})),
\end{equation}
where $D_{\HS}$ is the Hilbert-Schmidt distance and $\ZC^{\rm{A}}(\sigma^{\rm{A}})$ is the dephased (all off-diagonal elements set to zero) version of $\sigma^{\rm{A}}$. This quantity goes to zero if and only if $\mathcal F$ is consistent. For the partial trace case, we arrive at a similar cost function but with $\sigma^{\rm{A}}$ replaced by $\sigma^{\rm{SA}}$:
\begin{equation}
\label{eq:ptcost}
    C_{\rm{pt}}:= \sum_{\vec{\alpha}\ne\vec{\alpha}'} 
    \|\DC_{\rm{pt}}(\vec{\alpha},\vec{\alpha}')\|_{\HS}^2 =D_{\HS}(\sigma^{\rm{SA}}, \ZC^{\rm{A}}(\sigma^{\rm{SA}})).
\end{equation}
Here the notation $\ZC^{\rm{A}}(\sigma^{\rm{SA}})$ indicates that the dephasing operation only acts on system $\rm{A}$, and the absolute squares of Eq.~\eqref{eq:cost} have been generalized to Hilbert-Schmidt norms, $\|M\|_{\HS}^2:=\Tr(M\ad M)$. In the Methods section, we present quantum circuits that compute these cost functions from two copies of $\sigma^{\rm{A}}$ or $\sigma^{\rm{SA}}$. Derivations of the second equalities in Eq.~\eqref{eq:cost} and Eq.~\eqref{eq:ptcost} can be found in Appendix~\ref{sct:appC}. We remark that alternative cost functions may be useful, for example, to penalize families $\mathcal F$ with high entropy (see Methods) or to obtain a larger cost gradient by employing local instead of global observables (see Ref.~\cite{LaRose}).

The parameter optimization loop results in an approximately consistent family, $\mathcal F$, of histories, where the consistency parameter $\epsilon$ is upper bounded in terms of the final cost (see Methods). In Step \textbf{c} in Fig.~\ref{fig:flowchart}, we then generate the probabilities for the most likely histories by repeatedly preparing $\sigma^{\rm{A}}$ and measuring in the standard basis, where the measurement frequencies give the probabilities. (An alternative circuit that reads out any one of the exponentially many elements $\DC (\vec{\alpha},\vec{\alpha}')$ is introduced in Appendix~\ref{sct:appD}.)  Step \textbf{e}  shows how one prepares the set of projection operators for any given history in $\mathcal F$. These projectors can then be characterized with an efficient number of observables (i.e., avoiding full state tomography) to learn important information about the histories.

Let us discuss the scaling of VCH. With the potential exceptions of the Hamiltonian evolution and the projection operators, the complexity of our quantum circuits (i.e., the gate count, circuit depth, and total number of required qubits) scales linearly with both the system size $n$ and the number of times $k$ considered. The complexity of Hamiltonian evolution to some accuracy is problem dependent, but we typically expect polynomial scaling in $n$ for physical systems with properties like translational symmetry~\cite{Somma_Hamiltonian}. On the other hand, we consider the circuit depth for preparing the history projectors to be a refinement parameter. One can begin with a short-depth ansatz for the projectors and incrementally increase the depth to refine the ansatz, potentially improving the approximate consistency. We therefore expect the overall scaling of our quantum circuits to be polynomial in $n$ and $k$ for the anticipated use cases of VCH.

The complexity of minimizing our non-convex cost function is unknown, which is typical for VHQCAs. As classical methods for finding consistent families also involve optimizing over some parameterization for the projectors, classical methods also need to deal with this optimization complexity issue.

While the number of required repetitions of the probability readout step can scale inefficiently in $n$ and $k$ for certain families of histories, we assume that minimizing the cost outputs a family $\mathcal F$ for which the probability readout step is efficient. (See Methods for elaboration on this point.) 

This scaling behavior means that for systems that can be tractably simulated on a quantum computer and whose properties of interest are simple to implement, we achieve an exponential speedup and reduction in the needed resources as compared to classical approaches to this problem.

\subsection*{Experimental Implementations}

\begin{figure}[t]
\includegraphics[width=.84\columnwidth]{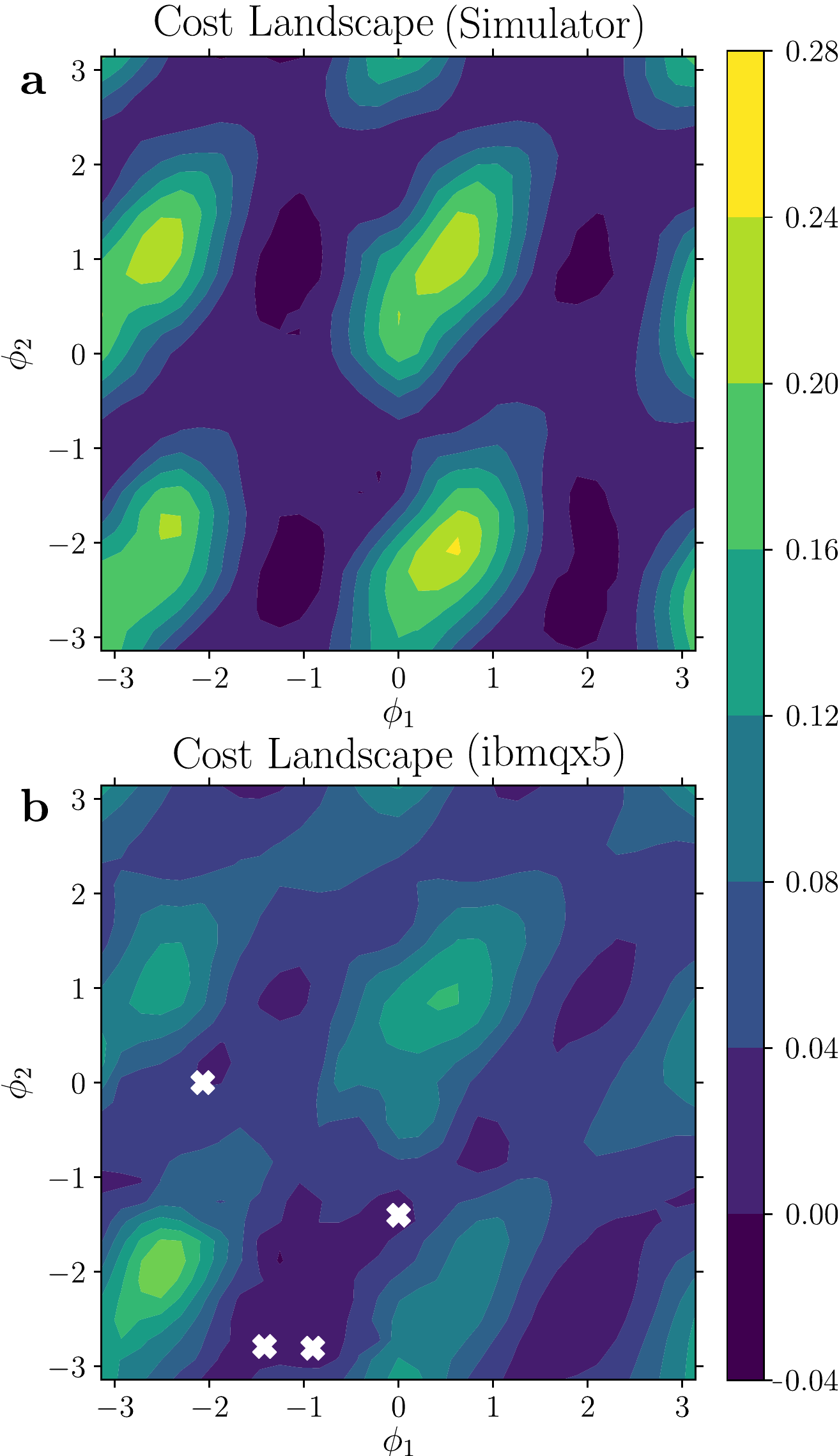}
\caption{Consistency of three-time histories for a spin-1/2 particle in a magnetic field, with initial state $\rho=\dya{+}$. The full-trace cost landscape, $C(\phi_1,\phi_2)$, is plotted as a function of the azimuths, $\phi_1$ and $\phi_2$, of the first and second projection bases, which we constrained to the $xy$ plane of the Bloch sphere. The point $(0,0)$ corresponds to both projections being along the $x$ axis. Consistency is expected everywhere along certain vertical lines ($\phi_1=2+n\pi\,\rm{rad}$), as they correspond to histories where the initial state is one of the projectors after the first time step, so there are no branches to interfere in the second time step. In addition, some slope-one lines ($\phi_2=\phi_1+(2+n\pi)\,\rm{rad}$) should be consistent, as they correspond to histories where the second projectors are the same as the first after time evolution, so no interference occurs in the second time step. Indeed, one can see valleys in the cost landscapes for these vertical and slope-one lines, when the cost is quantified on a simulator \textbf{a} and on the ibmqx5 quantum computer \textbf{b}. Note that negative cost values are possible due to finite statistics. The white ``x'' symbols in \textbf{b} mark some of the non-unique minima that the VCH algorithm found. }
\label{fig:B_landscapes}
\end{figure}

\emph{Spin in a magnetic field.} We now present an experimental demonstration of VCH on a cloud quantum computer. See Appendix~\ref{sct:appE} for further details on this implementation.
We examine the two time histories of a spin-1/2 particle in a magnetic field $B\hat{z}$, whose Hamiltonian is $H=-\gamma B \sigma_z$. The histories we consider have a time step $\Delta t$ between the initial state (chosen to be $\rho=\dya{+}$, with $|+\rangle=1/\sqrt{2}(|0\rangle+|1\rangle)$) and first projector, as well as between the first and second projector, chosen so that $\gamma B \Delta t = 2 \rm{rad}$. Additionally, we only consider projectors onto the $xy$ plane of the Bloch sphere, parameterized by their azimuth. For this model, Fig.~\ref{fig:B_landscapes} shows the landscape of the cost in Eq.~\eqref{eq:cost} for the ibmqx5 quantum processor~\cite{IBMQ16} as well as a simulator whose precision was limited by imposing the same finite statistics as were collected with the quantum processor. Several minima found by running VCH on ibmqx5 are superimposed on the landscape.  (All points found below a noise threshold were considered to be equally valid minima.) As these minima correspond reasonably well to theoretically consistent families, this represents a successful proof-of-principle implementation of VCH.
\begin{figure*}[ht]
    \centering
    \includegraphics[width=2\columnwidth]{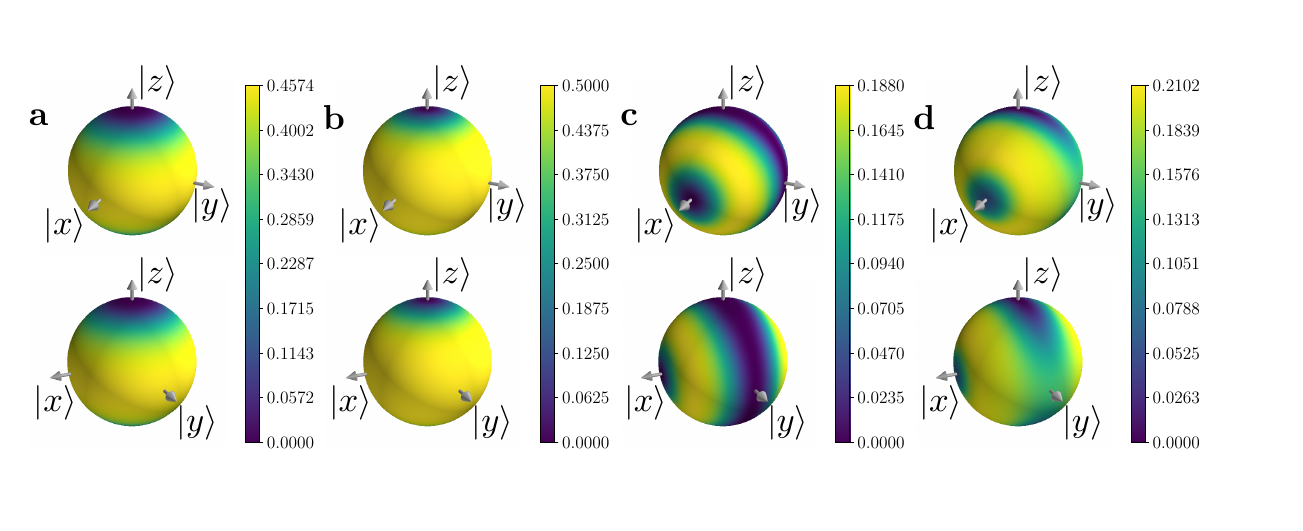}
    \caption{The cost landscape for stationary histories of the chiral molecule. Since the projectors in these stationary histories are always along a single axis, we plot the cost on points where this axis would intersect the surface of the Bloch sphere. The bottom row of spheres are the same as the top, but rotated for additional perspective. Panels \textbf{a} and \textbf{b} show the full and partial trace cost functions, respectively, for the case where the environment interactions are negligible ($\theta_z=5\,\rm{rad}$, $\theta_x=.01\,\rm{rad}$), and thus we find that the energy eigenbasis ($z$ axis) is the only consistent stationary family as all others will branch as they evolve. In contrast, panels \textbf{c} and \textbf{d} are the full and partial trace cost functions, respectively, for the case where the environment interactions dominate ($\theta_z=.01\,\rm{rad}$, $\theta_x=5\,\rm{rad}$). One can see in \textbf{c} and \textbf{d} a significant difference between the full and partial trace costs for the $y$ axis, meaning that this family of histories is consistent but not classical. In this regime, we also see that the chirality basis (the $x$ axis) is a local minimum for both cost functions and thus is approximately consistent and classical. For this chirality basis family, there is a $\sim 0.01\%$ chance that the molecule will change chirality during the evolution, showing that the quantum-to-classical transition leaves this system in a stabilized chiral state.}
    \label{fig:spheres}
\end{figure*}

\emph{Chiral molecule.} To highlight applications that will be possible on future hardware, we now turn to a simulated use of VCH to observe the quantum-to-classical transition for a chiral molecule~\cite{HundsP_Collisional_Stabilization,Patrick_chiral}. The chiral molecule has been modeled as a two level system where the right $\ket{R}$ and left $\ket{L}$ chirality states are described as $|R\rangle/|L\rangle=|+\rangle/|-\rangle=\frac{1}{\sqrt{2}}\left(|0\rangle \pm |1\rangle \right)$~\cite{Patrick_chiral}. A chiral molecule in isolation would tunnel between $\ket{R}$ and $\ket{L}$, but we consider the molecule to be in a gas, where collisions with other molecules convey information about the molecule's chirality to its environment. This information transfer is modeled by a rotation by angle $\theta_x$ about the $x$ axis of an environment qubit, controlled on the system's chirality, and for simplicity we suppose such collisions are evenly spaced at five points in time. (See Appendix~\ref{sct:appE} for further details.) We then consider simple families of  stationary histories \cite{Patrick_chiral}, where the projector set corresponds to the same basis at all five times (just after a collision occurs). Letting $\theta_z$ be the precession angle due to tunneling in the time between collisions, we can then explore the competition between decoherence and tunneling. Figure~\ref{fig:spheres} shows our results for this model. Notably we observe the transition from a quantum regime, where the chirality is not consistent, to a classical regime, where the chirality is both consistent and stable over time.

\section*{Discussion}
\label{sec:Discussion}

We expect VCH to revitalize interest in the CH approach to quantum mechanics by increasing its practical utility. Making it possible to apply the tools and concepts of quantum foundations to a wide array of physical situations, as VCH will, is an important step for our understanding of the physical world.  Specifically by providing an exponential speedup and reduction in resources over classical methods, VCH will provide a way to study phenomena including the quantum-to-classical transition~\cite{Paz_classicality,Finkelstein_pt,RZZ}, dynamics of quantum phase transitions~\cite{zurek2005dynamics}, quantum biological processes~\cite{lloyd_photosynthesis}, conformational changes~\cite{protien_folding}, and many other complex phenomena that so far have been computationally intractable. In addition, VCH could be applied to study quantum algorithms themselves~\cite{Poulin_CH_for_qip}. In order to highlight such potential applications and examine their resource requirements, we now focus on two of them: the emergence of classical diffusive dynamics in quantum spin systems and the appearance of defined pathways in protein folding.

In the context of Nuclear Magnetic Resonance (NMR) experiments, it has long been known that systems with many spins obey a classical diffusion equation while smaller spin systems undergo Rabi oscillations.  Despite the long history of spin diffusion studies~\cite{original_spin_diff,mean_extrapolation_spin_diff,thesis_spin_diff}, there is still no derivation of the transition from quantum oscillations to classical diffusion that can predict the size of the system where we should find that transition, or the nature of the transition. Applying VCH to the study of histories of spin systems would clarify this point by showing the scale and abruptness with which the diffusive behavior emerges. Since spin diffusion has been observed for systems as small as $\sim30,000$ spins~\cite{Quantum_dot_diffusion}, we estimate that between $\sim10^2$ and $\sim10^3$ qubits would allow us to study this transition. For more details about this application, see Appendix~\ref{sct:appF}.

In the protein folding community there are currently two main schools of thought on how proteins fold. The first is that proteins fold along well determined pathways with discrete folding units (foldons)~\cite{single-path_folding}, while the second is that there should be a funnel in the energy landscape of folding configurations, causing the system to explore a wide range of configurations before settling into the final state~\cite{multi-path_folding}. The deterministic pathways of the foldon model are favored by NMR experiments, raising the question of whether these views can be reconciled~\cite{single-path_folding}. By providing the means to study the dynamic emergence of classical paths, i.e., the quantum-to-classical transition for proteins, VCH could resolve this discrepancy. For this purpose, we estimate that between $\sim10^3$ to $\sim10^4$ qubits will be needed. See Appendix~\ref{sct:appF} for more details on this application and resource estimate.

Finally, our work highlights the synergy of two distinct fields, quantum foundations and quantum computational algorithms, and hopefully will inspire further research into their intersection.

\section*{Methods} 
\subsection*{Evaluation of the Cost}
\begin{figure}[h]
    \centering
    \includegraphics[width=\columnwidth]{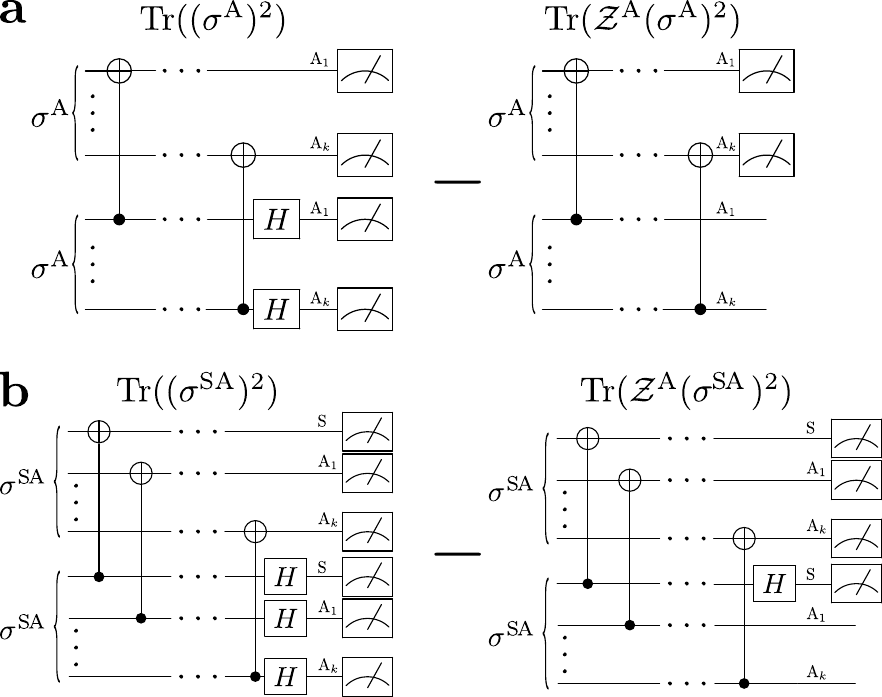}
    \caption{Circuits for computing the cost functions. Panel \textbf{a} shows the circuits for the full trace cost $C$ function and panel \textbf{b} shows the circuit for the partial trace cost $C_{\rm{pt}}$.}
    \label{fig:cost}
\end{figure}

Figure~\ref{fig:cost} shows the circuits for computing the full trace cost (partial trace cost) from two copies of $\sigma^{\rm{A}}$ ($\sigma^{\rm{SA}}$). Note that both costs can be written as a difference of purities:
\begin{align}
    C &=  \Tr ((\sigma^{\rm{A}})^2) -\Tr (\ZC^{\rm{A}} (\sigma^{\rm{A}})^2)\\
    C_{\rm{pt}} &=  \Tr ((\sigma^{\rm{SA}})^2) -\Tr (\ZC^{\rm{A}} (\sigma^{\rm{SA}})^2)\,.
\end{align}
The $\Tr ((\sigma^{\rm{A}})^2)$ and $\Tr ((\sigma^{\rm{SA}})^2)$ terms are computed via the Swap Test, with a depth-two circuit and classical post-processing that scales linearly in the number of qubits~\cite{garcia2013swap,
cincio2018learning}. A similar but even simpler circuit, called the Diagonalized Inner Product (DIP) Test~\cite{LaRose}, calculates the $\Tr (\ZC^{\rm{A}} (\sigma^{\rm{A}})^2)$ term with a depth one circuit and no post-processing. Finally, the $\Tr (\ZC^{\rm{A}} (\sigma^{\rm{SA}})^2)$ term is evaluated with the Partial-DIP (PDIP) Test~\cite{LaRose}, a depth-two circuit that is a hybridization of the Swap Test and the DIP Test.

\subsection*{Precision of probability readout}

One does not know \textit{a priori} how many histories will be characterized in the probability readout step (Fig.~\ref{fig:flowchart}\textbf{c}). Due to statistical noise, the probability of histories with greater probability will be determined with greater relative precision than those with lesser probability. Hence, it is reasonable to set a precision (or statistical noise) threshold, $\epsilon$. Let $N_{\text{readout}}$ be the number of repetitions of the probability readout circuit. Then, histories $\YC^{\vec{\alpha}}$ whose bitstring $\vec{\alpha}$ occurs with frequency $f_{\vec{\alpha}} <  \sqrt{N_{\text{readout}}} / \epsilon_{\max}$ should be ignored, since their probabilities $p(\vec{\alpha})=f_{\vec{\alpha}}/N_{\text{readout}}$ were not characterized with the desired precision. We separate $\mathcal F$ into the set $\mathcal F_c$ of histories whose probabilities are above the precision threshold (which we previously referred to loosely as the most likely histories), and the set of all other histories in $\mathcal F$:
\begin{equation}
\label{eq:split family}
    \mathcal F=\mathcal F_c \cup \overline{\mathcal F_c}.
\end{equation}

Computational complexity can be hidden in the value of $N_{\text{readout}}$ needed to obtain a desired precision for the probabilities of histories of interest. This issue is closely connected to the entropy of the set $\{\DC(\vec{\alpha}, \vec{\alpha})\}$, or equivalently, the entropy of the quantum state $\ZC^{\rm{A}}(\sigma^{\rm{A}})$. When $\ZC^{\rm{A}}(\sigma^{\rm{A}})$ is high entropy, an exponentially large number of histories may have non-zero probability, and hence $N_{\text{readout}}$ would need to grow exponentially. VCH is therefore better suited to applications where there is a small subset of the histories that are far more probable than the rest. In the parameter optimization loop, one can select for families with this property by penalizing families for which $\ZC^{\rm{A}}(\sigma^{\rm{A}})$ has high entropy. Specifically, by noting that $P := \Tr(\ZC^{\rm{A}}(\sigma^{\rm{A}})^2)$ can be efficiently computed via the circuit in Fig.~\ref{fig:cost}\textbf{a}, one can modify the costs functions in Eq.~\eqref{eq:cost} and Eq.~\eqref{eq:ptcost} to be $\tilde{C} = C / P$ and $\tilde{C}_{\rm{pt}} = C_{\rm{pt}} / P$. 

We remark that classicality is intimately connected to predictability, with the emergence of classicality linked to the so-called predictability sieve~\cite{Zurek_predictability,Dalvit_predictability}. Since the CH formalism is typically used to find classical families, this implies predictable families (i.e., families with low entropy or high purity $P$) are arguably of the most interest. Hence, our modified cost function $\tilde{C}$ also serves to select those consistent families with histories that are the most predictable, and therefore the most classical.

\subsection*{Approximate Consistency}

Here we discuss how VCH outputs an upper bound on the consistency parameter $\epsilon$. Let us first relate the cost $C$ to $\epsilon$. For any pair of histories $\YC^{\vec{\alpha}}$ and $\YC^{\vec{\alpha'}}$ in $\mathcal F$,
\begin{equation}
\label{eq:off diagonal bound}
    |\DC(\vec{\alpha},\vec{\alpha}')|^2\leq C/2,
\end{equation}
which follows from Eq.~\eqref{eq:cost} and the fact that $|\DC(\vec{\alpha},\vec{\alpha}')|=|\DC(\vec{\alpha}',\vec{\alpha})|$. Let us define
\begin{equation}
    \label{eq:C to eps}
     \epsilon_{\vec{\alpha},\vec{\alpha}'}:= \sqrt{\frac{C}{2\DC(\vec{\alpha},\vec{\alpha})\DC(\vec{\alpha}',\vec{\alpha}')}}.
\end{equation}
Then it follows from Eq.~\eqref{eq:off diagonal bound} that
\begin{equation}
\label{eq:GGH approx consistency conditionv2}
    |\DC(\vec{\alpha},\vec{\alpha}')|^2 \leq \epsilon_{\vec{\alpha},\vec{\alpha}'}^2\DC(\vec{\alpha},\vec{\alpha})\DC(\vec{\alpha}',\vec{\alpha}'),
\end{equation}
which corresponds to the approximate consistency condition from Eq.~\eqref{eq:GGH approx consistency condition}. Hence, probablity sum rules for these two histories are satisfied within error $\epsilon_{\vec{\alpha},\vec{\alpha}'}$, which can be calculated from Eq.~\eqref{eq:C to eps} for histories in $\mathcal F_c$ since the probabilites are known for these histories.

Next, consider histories in $\overline{\mathcal F_c}$. As we do not have enough information to differentiate these histories, we advocate combining the elements of $\overline{\mathcal F_c}$ into a single coarse-grained history $\YC^{\vec{\gamma}}$. 

Let $\YC^{\vec{\beta}}$ be the least likely history in $\mathcal F_c$. Then defining $\delta^2=\DC(\vec{\gamma},\vec{\gamma})/ \DC(\vec{\beta},\vec{\beta})$, we can make use of the positive semi-definite property of $\sigma^{\rm{A}}$ to write:
\begin{equation}
    |\DC(\vec{\gamma},\vec{\beta})|^2 \leq \DC(\vec{\gamma},\vec{\gamma})\DC(\vec{\beta},\vec{\beta}) = \delta^2 \DC(\vec{\beta},\vec{\beta})^2. 
\end{equation}
Since $\YC^{\vec{\beta}}$ is the least likely history in $\mathcal F_c$, this expression then lets us bound the error on the probability sum rule (giving a weaker approximate consistency condition~\cite{approximate_consistency}) between $\YC^{\vec{\gamma}}$ and any $\YC^{\vec{\alpha}}\in \mathcal F_c$ as:
\begin{eqnarray}
    |\DC(\vec{\gamma},\vec{\alpha})|&\leq&\delta \DC(\vec{\alpha},\vec{\alpha})\nonumber\\&\leq&\delta(\DC(\vec{\gamma},\vec{\gamma})+\DC(\vec{\alpha},\vec{\alpha}))\,.
\end{eqnarray}

It is then possible to characterize the approximate consistency of the histories of $\mathcal F$ pairwise with $\epsilon_{\vec{\alpha},\vec{\alpha}'}$ and $\delta$. Alternatively, to give an upper bound on the overall consistency $\epsilon$, we take the greatest of these pairwise bounds:
\begin{equation}
    \epsilon\leq \rm{max}(\{\epsilon_{\vec{\alpha},\vec{\alpha}'}\}\cup \{\delta\}).
\end{equation}

For those applications where we are working with the partial trace consistency, the notion of approximate consistency is somewhat more obscured. In order to generate probabilities and bound $\epsilon$, we therefore recommend evaluating the full trace cost function at the minimum found with the partial trace cost. This approach is helpful since any partial trace consistent family will also be full trace consistent and the partial trace consistency does not directly allow one to discuss probabilities in the same way. Taking this approach allows us to then directly utilize the approximate consistency framework above.

\begin{acknowledgements}

We thank IBM for the use of their quantum processor. The views expressed in this article are those of the authors and not of IBM.
This work was supported by the U.S. Department of Energy (DOE), Office of Science, Office of High Energy Physics, QuantISED program, and also by the U.S. DOE, Office of Science, Basic Energy Sciences, Materials Sciences and Engineering Division, Condensed Matter Theory Program.  All authors acknowledge support from the LDRD program at Los Alamos National Laboratory (LANL).  LC was also supported by the DOE through the J. Robert Oppenheimer fellowship. ATS and PJC additionally acknowledge support from the LANL ASC Beyond Moore's Law project. Finally, WHZ acknowledges partial support by the Foundational Questions Institute grant FQXi-1821, and Franklin Fetzer Fund, a donor advised fund of the Silicon Valley Community Foundation.
\end{acknowledgements}
\section*{Author Contributions}
All authors contributed to the preparation and revision of the manuscript. P.J.C. invented the algorithm and developed the basic formalism. A.A. designed and carried out the experimental implementations, analyzed the results, and contributed to the formalism. L.C., A.T.S., and W.H.Z. consulted on all stages of the project.
\section*{Competing Interests}
The authors declare no competing interests.
\section*{Data Availability}
The data used to create the figures in this article are available upon request. Requests should be sent to the corresponding author.

\clearpage

\appendix
\setcounter{page}{1}
\renewcommand\thefigure{S.\arabic{figure}}
\setcounter{figure}{0}

\onecolumngrid
\begin{center}
\Large{ Supplementary Material for \\ ``Variational Consistent Histories as a Hybrid Algorithm for Quantum Foundations''}
\end{center}
\twocolumngrid

\section{Generalizations}\label{sct:appA}

Here we discuss various generalizations of the circuits shown in the main text, which presented our VCH algorithm for the special case of branch-independent histories of a one-qubit system $S$ with no environment $E$.

\subsection*{Multi-Qubit Systems}

The circuits in the main text showed systems $S$ composed of a single qubit. The generalization to multi-qubit systems is straightforward. We must discuss the generalizations of both the state preparation circuit in Fig.~\ref{fig:flowchart} as well as the cost evaluation circuits in Fig.~\ref{fig:cost}.

\begin{figure}[h]
    \centering
    \includegraphics{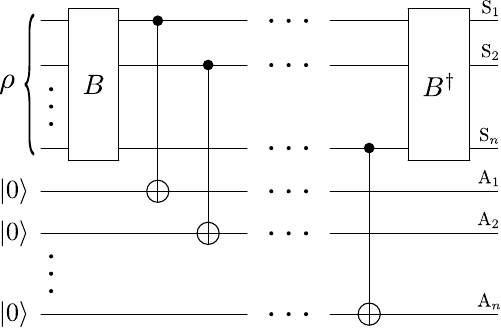}
    \caption{The generalization of our state preparation circuit to multi-qubit systems $S$. In this example, we show the portion of the circuit that entangles the system and the ancillas, for the special case of a fine-grained set of projectors. In this fine-grained case, one employs the same number of ancilla qubits as are in $S$, i.e., $n$ qubits.}
    \label{fig:big_example}
\end{figure}

Figure~\ref{fig:big_example} illustrates how the state preparation circuit generalizes to multi-qubit systems. In particular, this figure shows how a portion of state preparation circuit (the portion that entangles the system to the ancillas) generalizes for the case of a fine-grained set of projectors. (Note that the case of a coarse-grained set of projectors is discussed in the next subsection.)

The cost evaluation circuits in Fig.~\ref{fig:cost} generalize as follows. For fine-grained histories, one needs $n$ ancillas for each time step and hence a total of $nk$ ancillas. The circuits in Fig.~\ref{fig:cost} shown for $k$ ancillas generalize in a straightforward way, where now one has $nk$ ancilla systems. In addition, the circuits in Fig.~\ref{fig:cost}\textbf{b} also involve the $S$ system, and hence all $n$ qubits in $S$ must be included in this circuit. Again, these $n$ qubits are included in the most straightforward way (in the same way that the single qubit $S$ system appears in the circuits in Fig.~\ref{fig:cost}\textbf{b}).

\subsection*{Coarse Grained Histories}

Multi-qubit systems $S$ allow for non-trivial coarse-grained histories.
In such families of histories, the sets $P_j$ are composed of projectors whose ranks are possibly greater than one. We remark that coarse-grained histories are often important to the study of macroscropic systems and the quantum-to-classical transition. VCH can easily be adapted to study coarse-grained histories as follows.

For each time $t_j$, one should decide (prior to running VCH) projector ranks that one is interested in. VCH will then optimize over sets of projectors with these particular ranks. The projector ranks can therefore be viewed as hyperparameters, i.e., parameters that one fixes for a given run of VCH.

\begin{figure}[h]
    \centering
    \includegraphics[width=\columnwidth]{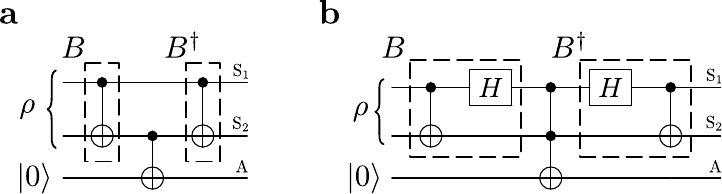}
    \caption{Examples of implementing coarse-grained projector sets in our state preparation circuit, when $S$ corresponds to two spin-1/2 particles. The projectors in \textbf{a} record whether the two spins are aligned or anti-aligned, while the projectors in \textbf{b} differentiate between the spin singlet and spin triplet states.}
    \label{fig:align_circ}
\end{figure}

For instance, suppose $S$ is composed of a pair of spins. In this case, Fig.~\ref{fig:align_circ} shows two examples of the state preparation circuit for a single time step. In the first example, Fig.~\ref{fig:align_circ}\textbf{a}, we consider a projector set that contains two rank-two projectors revealing whether the spins were aligned or anti-aligned. In the second example, Fig.~\ref{fig:align_circ}\textbf{b}, we consider a projector set that contains a rank-three and a rank-one projector that respectively indicate whether the spins are in the triplet states or the the singlet state. Note that the ranks of the projectors are determined by the gate that entangles the system to the ancilla, which is a single CNOT gate in Fig.~\ref{fig:align_circ}\textbf{a} and a Toffoli gate in Fig.~\ref{fig:align_circ}\textbf{b}. Hence the choice of the projector ranks (mentioned in the previous paragraph) translates into a choice of gate sequence that entangles the system to the ancilla.

\subsection*{Nontrivial Environments}

\begin{figure}[h]
    \centering
    \includegraphics{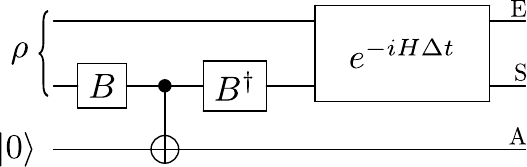}
    \caption{Simple example with an environment $\rm{E}$. The projectors still only act on $\rm{S}$, but the evolution includes both $\rm{S}$ and $\rm{E}$.}
    \label{fig:env}
\end{figure}

For many applications of VCH, (e.g., the chiral molecule example in the main text) it will be helpful to explicitly model an environment $\rm{E}$. We can think of this case as a particular choice of coarse graining where the projectors we consider only act on a subsystem of our model (the $S$ system) and do not directly record any information about $\rm{E}$. Note that the Hamiltonian evolution involves both $\rm{S}$ and $\rm{E}$, as shown in Fig.~\ref{fig:env}. 

\subsection*{Branch Dependent Histories}

A final generalization that we consider are families of branch dependent histories~\cite{branch-dependent}, or histories where the projector set at a given time may depend on the properties of the system at earlier points in the histories. VCH can accommodate these histories, as follows.

The basic idea is that the unitary gate $B_j$ that determines the projector set at time $t_j$ now becomes a controlled unitary. Specifically, the control system(s) for $B_j$ are (potentially) all the ancilla qubits associated with times $t_i < t_j$. So the choice of projector set at some time is influenced by the ancilla states for earlier times.

\begin{figure}[h]
    \centering
    \includegraphics[width=\columnwidth]{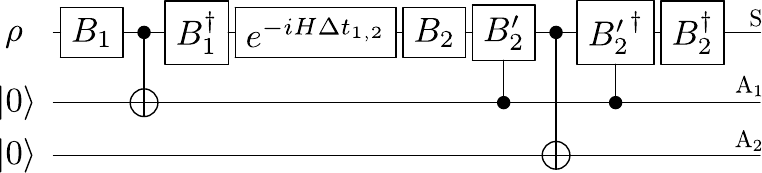}
    \caption{Example implementation of a branch dependent projector set in our state preparation circuit. In this circuit, depending upon the result for $t_1$, either $B_2$ or the product $B_2'B_2$ defines the projector set for the second time.}
    \label{fig:branch_circ}
\end{figure}

Figure~\ref{fig:branch_circ} shows an example of what this looks like, for the special case of only two times. In this figure, if the first ancilla is in the $\ket{0}$ state ($\ket{1}$ state), then the $B_2$ unitary ($B_2' B_2$ unitary) is applied at the second time step. For more general cases, the $B_2'$ unitary shown here would be replaced by a sequence of controlled unitaries controlled by different ancilla qubits.

\section{Generalized state preparation} \label{sct:appB}

\begin{figure*}[ht]
    \centering
    \includegraphics[width=\textwidth]{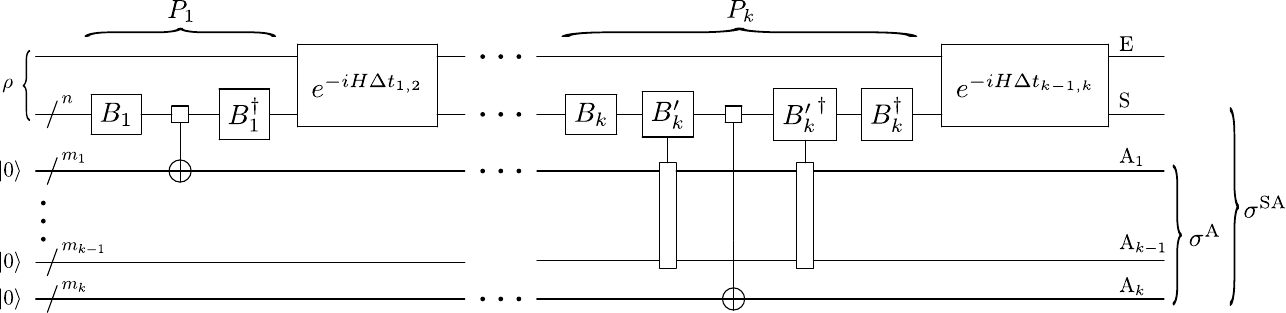}
    \caption{The generalized state preparation circuit. A similar circuit is included as part of the flowchart, but this version incorporates larger systems and branch dependence explicitly. The multiqubit gate \raisebox{-0.2cm}{\protect\includegraphics[height=0.6cm]{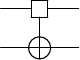}}
    denotes a set of entangling gates controlled on the standard basis of the system qubits, while
    \raisebox{-0.2cm}{\protect\includegraphics[height=0.6cm]{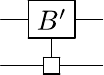}}
    represents a parameterized unitary acting on the system controlled on the standard basis of the ancilla qubits.
    }
    \label{fig:gen_state_prep}
\end{figure*}

We now present the details of our generalized state preparation circuit (as shown in Fig.~\ref{fig:gen_state_prep}) and show that $\sigma^{\rm{SA}}$ and $\sigma^{\rm{A}}$ have the properties we claim in the main text. Note that our treatment here includes all of the generalizations discussed above in Appendix~\ref{sct:appA}. We begin with the input state $\rho^{\rm{SE}}\otimes|\vec{0}\>\<\vec{0}|^{\rm{A}}$ (where the superscript $\rm{SE}$ denotes the system and its environment and $\rm{A}$ denotes the ancillas). We then apply the gate sequence associated with the $P_1$ projector set, which includes $B_1$, a multi-qubit gate that entangles $S$ and $A$ (which we refer to as the ``entangling gate''), and then $B_1^\dagger$. This gives the state:
\begin{eqnarray}
\sum_{\alpha_1,\alpha_1'}&&\Big[P_1^{\alpha_1}\rho^{\rm{SE}}P_1^{\alpha_1'\dagger}\Big]\otimes\Big[|\alpha_1\>\<\alpha_1'|\otimes|\vec{0}\>\<\vec{0}|\Big]^{\rm{A}}.
\end{eqnarray}
Note that the system and ancilla are (possibly) entangled at this point.
 
Next in our state preparation circuit is the time evolution from $t_1$ to $t_2$, given by $e^{-i H \Delta t_{1,2}}$. This is followed by the gate sequence associated with $P_2$, which in general may be branch dependent. The resulting state is
\begin{eqnarray}
    \sum_{\alpha_1,\alpha_1',\alpha_2,\alpha_2'}&&\Big[P_2^{\alpha_2}({\alpha_1})e^{-i H \Delta t_{1,2}} P_1^{\alpha_1}\rho^{\rm{SE}}P_1^{\alpha_1'\dagger}e^{i H \Delta t_{1,2}}\nonumber \\
    &&P_2^{\alpha_2'\dagger}({\alpha_1})\Big]\otimes\Big[\dyad{\alpha_1}{\alpha_1 '}\otimes \dyad{\alpha_2}{\alpha_2 '} \nonumber\\
    &&\otimes \dyad{\vec{0}}{\vec{0}}\Big]^{\rm{A}},
\end{eqnarray}
where the notation $P_2^{\alpha_2}({\alpha_1})$ indicates that the second projector set depends on $\alpha_1$. Repeating this state evolution until we have applied the gate sequences associated with all $k$ projector sets (and switching to the Heisenberg picture), we end up with
\begin{eqnarray}
    \sum_{\vec{\alpha},\vec{\alpha}'}&&\Big[P_{\rm{k}}^{\alpha_{\rm{k}}}(t_{\rm{k}})\dots P_2^{\alpha_2}(t_2) P_1^{\alpha_1}(t_1)\rho^{\rm{SE}}\nonumber \\
    &&\quad P_1^{\alpha_1'}(t_1)^{\dagger}P_2^{\alpha_2'}(t_2)^{\dagger}\dots P_{\rm{k}}^{\alpha_{\rm{k}}'}(t_{\rm{k}})^{\dagger}\Big]\nonumber \\
    &&\otimes\Big[(|\alpha_1\>\<\alpha_1'|)\otimes(|\alpha_2\>\<\alpha_2'|)\otimes\dots\otimes(|\alpha_{\rm{k}}\>\<\alpha_{\rm{k}}'|)\Big]^{\rm{A}}\nonumber \\
    =\sum_{\vec{\alpha},\vec{\alpha}'}&&\mathcal{C}^{\vec{\alpha}}\rho^{\rm{SE}} \mathcal{C}^{\vec{\alpha}'\dagger}\otimes(|\vec{\alpha}\>\<\vec{\alpha}'|)^{\rm{A}}
\end{eqnarray}
Note that we have suppressed explicit branch dependence here to simplify notation. Branch dependence does not alter the formalism except to make the later projectors functions of the earlier $\alpha_i$'s, so our treatment remains fully general.

If we then trace out the environment (which in the circuit means not measuring it) we are then left with $\sigma^{\rm{SA}}$:
\begin{equation}
    \label{eq:trace_E-sigmaSA}
   \sigma^{\rm{SA}} =\sum_{\vec{\alpha},\vec{\alpha}'}\Tr_{\rm{E}}(\mathcal{C}^{\vec{\alpha}}\rho^{\rm{SE}} \mathcal{C}^{\vec{\alpha}'\dagger})\otimes(|\vec{\alpha}\>\<\vec{\alpha}'|)^{\rm{A}}.
\end{equation}
By examining Eq.~\eqref{eq:trace_E-sigmaSA}, we can see that $(\mathbb{1}\otimes\<\vec{\alpha}|)\sigma^{\rm{SA}}(\mathbb{1}\otimes|\vec{\alpha}\>)$ is precisely $\DC_{\rm{pt}}(\vec{\alpha},\vec{\alpha}')=\Tr_{\rm{E}}(\mathcal{C}^{\vec{\alpha}}\rho^{\rm{SE}}\, \mathcal{C}^{\vec{\alpha}'\dagger})$. Further, if we similarly trace over the system $\rm{S}$, we get:
\begin{equation}
    \label{eq:trace_SE-sigmaA}
   \sigma^{\rm{A}} =\sum_{\vec{\alpha},\vec{\alpha}'}\Tr(\mathcal{C}^{\vec{\alpha}}\rho^{\rm{SE}} \mathcal{C}^{\vec{\alpha}'\dagger})(|\vec{\alpha}\>\<\vec{\alpha}'|)^{\rm{A}}.
\end{equation}
We can thus see that we have prepared a density matrix whose elements are $\mathcal{D}(\vec{\alpha},\vec{\alpha}')=\Tr(\mathcal{C}^{\vec{\alpha}}\rho^{\rm{SE}}\mathcal{C}^{\vec{\alpha}'\dagger})$, as claimed in the main text.

\section{Derivation of Cost Functions} \label{sct:appC}

\subsection*{Full trace cost}

Let us now derive the equivalence stated in the definition of our full trace cost function, Eq.~\eqref{eq:cost}. Starting with the definition of ${C}$ we have:
\begin{eqnarray}
    {C}&:= &\sum_{\vec{\alpha}\ne\vec{\alpha}'}|\DC(\vec{\alpha},\vec{\alpha}')|^2\nonumber\\
    &=&\sum_{\vec{\alpha}\ne\vec{\alpha}'}\<\vec{\alpha}|\sigma^{\rm{A}}|\vec{\alpha}'\>\<\vec{\alpha}'|\sigma^{\rm{A}}|\vec{\alpha}\> \nonumber\\
    &=& \sum_{\vec{\alpha}\ne\vec{\alpha}'}\Tr \left( (|\vec{\alpha}\>\<\vec{\alpha}|)\sigma^{\rm{A}}(|\vec{\alpha}'\>\<\vec{\alpha}'|)\sigma^{\rm{A}}\right)\nonumber\\
    &=& \sum_{\vec{\alpha},\vec{\alpha}'}\Tr \left( (|\vec{\alpha}\>\<\vec{\alpha}|)\sigma^{\rm{A}}(|\vec{\alpha}'\>\<\vec{\alpha}'|)\sigma^{\rm{A}}\right)\nonumber \\
    &&-\sum_{\vec{\alpha}} \Tr \left( (|\vec{\alpha}\>\<\vec{\alpha}|)\sigma^{\rm{A}}(|\vec{\alpha}\>\<\vec{\alpha}|)\sigma^{\rm{A}}\right) \nonumber\\
    &=&\Tr ((\sigma^{\rm{A}})^2)-\Tr (\ZC^{\rm{A}} (\sigma^{\rm{A}})^2) \nonumber\\
    &=&\rm{D}_{\rm{HS}}(\sigma^{\rm{A}},\ZC^{\rm{A}}(\sigma^{\rm{A}})).
\end{eqnarray}
Therefore, the circuits we use to calculate $\Tr ((\sigma^{\rm{A}})^2)$ and $\Tr (\ZC^{\rm{A}} (\sigma^{\rm{A}})^2)$ implement this cost function as claimed.

\subsection*{Partial trace cost}

Arriving at the expression for the partial trace cost function (Eq.~\eqref{eq:ptcost}) is similar if slightly more complicated:
\begin{eqnarray}
    \rm{C_{\rm{pt}}}&:=&\sum_{\vec{\alpha}\ne\vec{\alpha}'}\|\DC_{\rm{pt}}(\vec{\alpha},\vec{\alpha}')\|_{\rm{HS}}^2 \nonumber\\
    &=& \sum_{\vec{\alpha}\ne\vec{\alpha}'}\Tr_{\rm{S}} \left(\DC_{\rm{pt}}(\vec{\alpha},\vec{\alpha}')\DC_{\rm{pt}}(\vec{\alpha},\vec{\alpha}')^\dagger \right)\nonumber\\
    &=& \sum_{\vec{\alpha}\ne\vec{\alpha}'}\Tr_{\rm{S}} \left((\mathbb{1}\otimes\<\vec{\alpha}|) (\mathbb{1}\otimes|\vec{\alpha}\>\<\vec{\alpha}|)\sigma^{\rm{SA}}\right.\nonumber \\
    &&\qquad\qquad\,\, \left. (\mathbb{1}\otimes|\vec{\alpha}'\>\<\vec{\alpha}'|)\sigma^{\rm{SA}}(\mathbb{1}\otimes|\vec{\alpha}\>) \right)\nonumber\\
    &=& \sum_{\vec{\alpha}\ne\vec{\alpha}'}\Tr \left( (\mathbb{1}\otimes|\vec{\alpha}\>\<\vec{\alpha}|)\sigma^{\rm{SA}}(\mathbb{1}\otimes|\vec{\alpha}'\>\<\vec{\alpha}'|)\sigma^{\rm{SA}} \right)\nonumber\\
    &=& \sum_{\vec{\alpha},\vec{\alpha}'}\Tr \left( (\mathbb{1}\otimes|\vec{\alpha}\>\<\vec{\alpha}|)\sigma^{\rm{SA}}(\mathbb{1}\otimes|\vec{\alpha}'\>\<\vec{\alpha}'|)\sigma^{\rm{SA}}\right)\nonumber \\
    &&-\sum_{\vec{\alpha}} \Tr \left( (\mathbb{1}\otimes|\vec{\alpha}\>\<\vec{\alpha}|)\sigma^{\rm{SA}}(\mathbb{1}\otimes|\vec{\alpha}\>\<\vec{\alpha}|)\sigma^{\rm{SA}}\right) \nonumber\\
    &=& \Tr ((\sigma^{\rm{SA}})^2)-\Tr (\ZC^{\rm{A}}(\sigma^{\rm{SA}})^2) \nonumber\\
    &=&\rm{D}_{\rm{HS}}(\sigma^{\rm{SA}},\ZC^{\rm{A}}(\sigma^{\rm{SA}})).
\end{eqnarray}
As with the full trace cost function, the circuits we use to calculate $\Tr ((\sigma^{\rm{SA}})^2)$ and $\Tr (\ZC^{\rm{A}} (\sigma^{\rm{SA}})^2)$ thus implement this cost function as claimed.
\section{Reading out the Decoherence Functional Elements} \label{sct:appD}
\begin{figure}
    \centering
    \includegraphics[width=\columnwidth]{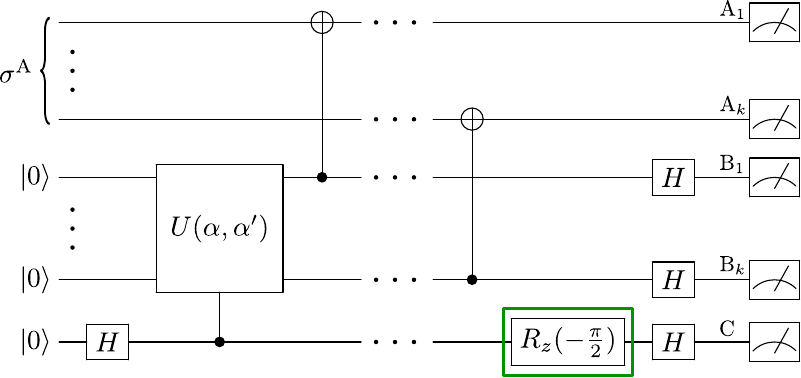}
    \caption{Circuit to read out $\DC(\vec{\alpha},\vec{\alpha}')$. The controlled $U(\vec{\alpha},\vec{\alpha}')$ prepares the state $|\alpha\>$ on the $\rm{B}$ registers when the control qubit is in the state $|0\>$  and $|\alpha'\>$ when the control qubit is in the state $|1\>$, so the combination of the Hadamard gate on $\rm{C}$ and the controlled $U(\vec{\alpha},\vec{\alpha}')$ prepares a superposition of the histories. The $z$-rotation in the green box is excluded when we calculate the real part of $\DC(\vec{\alpha},\vec{\alpha}')$ and included when we calculate the imaginary part. The post processing is described in the text.}
    \label{fig:check_circ}
\end{figure}
While VCH avoids the need to compute the exponentially many $\DC(\vec{\alpha},\vec{\alpha}')$'s in order to determine the consistency of a family $\mathcal F$, we do have the ability to efficiently read out any particular $\DC(\vec{\alpha},\vec{\alpha}')$ if desired. Figure~\ref{fig:check_circ} shows the circuit that one can use to read the real and/or imaginary parts of $\DC(\vec{\alpha},\vec{\alpha}')$ out for $\vec{\alpha}\ne\vec{\alpha}'$. The post-processing is similar to that of the Swap test~\cite{garcia2013swap,
cincio2018learning}, except that we add a conditional statement. 

When we exclude the $z$-rotation, conditioned on the control qubit $\rm{C}$ being measured in the state $|0\>$ we perform the Swap test between the $\rm{A}$ and $\rm{B}$ registers to get:
\begin{eqnarray}
    R_0&=&\Tr \left(\sigma^{\rm{A}}\left[ \frac{1}{2}(|\vec{\alpha}\>+|\vec{\alpha}'\>)(\<\vec{\alpha}|+\<\vec{\alpha}'|)\right]\right)\nonumber \\
    &=& \frac{1}{2}\left(\DC(\vec{\alpha},\vec{\alpha})+\DC(\vec{\alpha}',\vec{\alpha}')+\DC(\vec{\alpha},\vec{\alpha}')+\DC(\vec{\alpha}',\vec{\alpha})\right)\nonumber\\
    &=& \frac{1}{2}\left(\DC(\vec{\alpha},\vec{\alpha})+\DC(\vec{\alpha}',\vec{\alpha}')\right)+\rm{Re}(\DC(\vec{\alpha},\vec{\alpha}')). 
\end{eqnarray}
If we instead condition on $\rm{C}$ being measured in the state $|1\>$ we perform the Swap test between the $\rm{A}$ and $\rm{B}$ registers to get:
\begin{eqnarray}
    R_1&=&\Tr \left(\sigma^{\rm{A}}\left[ \frac{1}{2}(|\vec{\alpha}\>-|\vec{\alpha}'\>)(\<\vec{\alpha}|-\<\vec{\alpha}'|)\right]\right)\nonumber \\
    &=& \frac{1}{2}\left(\DC(\vec{\alpha},\vec{\alpha})+\DC(\vec{\alpha}',\vec{\alpha}')\right)- \rm{Re}(\DC(\vec{\alpha},\vec{\alpha}')). 
\end{eqnarray}
Our method therefore separates the output based on the result of measuring $\rm{C}$, and then performs the usual Swap test post processing on each partition of the output counts to get $R_0$ and $R_1$. Finally, we combine these to get:
\begin{equation}
    \rm{Re}(\DC(\vec{\alpha},\vec{\alpha}'))=\frac{1}{2} (R_0-R_1) \; .
\end{equation}

Instead including that $z$-rotation gives us 
\begin{eqnarray}
    I_0&=&\Tr \left(\sigma^{\rm{A}}\left[ \frac{1}{2}(|\vec{\alpha}\>+i|\vec{\alpha}'\>)(\<\vec{\alpha}|-i\<\vec{\alpha}'|)\right]\right)\nonumber \\
    &=& \frac{1}{2}\left(\DC(\vec{\alpha},\vec{\alpha})+\DC(\vec{\alpha}',\vec{\alpha}')-i\DC(\vec{\alpha},\vec{\alpha}')+i\DC(\vec{\alpha}',\vec{\alpha})\right)\nonumber\\
    &=& \frac{1}{2}\left(\DC(\vec{\alpha},\vec{\alpha})+\DC(\vec{\alpha}',\vec{\alpha}')\right)-\rm{Im}(\DC(\vec{\alpha},\vec{\alpha}')),
\end{eqnarray}
conditioned on $\rm{C}$ being measured in the state $|0\>$. Similarly, conditioned on $\rm{C}$ being measured in the state $|1\>$ we find:
\begin{eqnarray}
    I_1&=&\Tr \left(\sigma^{\rm{A}}\left[ \frac{1}{2}(|\vec{\alpha}\>-i|\vec{\alpha}'\>)(\<\vec{\alpha}|+i\<\vec{\alpha}'|)\right]\right)\nonumber \\
    &=& \frac{1}{2}\left(\DC(\vec{\alpha},\vec{\alpha})+\DC(\vec{\alpha}',\vec{\alpha}')+i\DC(\vec{\alpha},\vec{\alpha}')-i\DC(\vec{\alpha}',\vec{\alpha})\right)\nonumber\\
    &=& \frac{1}{2}\left(\DC(\vec{\alpha},\vec{\alpha})+\DC(\vec{\alpha}',\vec{\alpha}')\right)+\rm{Im}(\DC(\vec{\alpha},\vec{\alpha}')). 
\end{eqnarray}
Again, we combine these to get:
\begin{equation}
    \rm{Im}(\DC(\vec{\alpha},\vec{\alpha}'))=\frac{1}{2} (I_1-I_0)
\end{equation}

We also note that the controlled $U(\vec{\alpha},\vec{\alpha}')$ we have made use of here can be implemented with depth that scales linearly in the number of bits by which $|\vec{\alpha}\>$ and $|\vec{\alpha}'\>$ differ. This is accomplished by acting with $X$ gates on all of the registers where the bit-string associated with $|\vec{\alpha}\>$ is 1 followed by CNOT gates from $\rm{C}$ to each of the registers where the bit-strings for $|\vec{\alpha}\>$ and $|\vec{\alpha}'\>$ differ.

Finally, we comment that reading out  $\DC(\vec{\alpha},\vec{\alpha})$ is simpler than the general case as we merely have to prepare $|\vec{\alpha}\>\<\vec{\alpha}|$ (which consists of a single layer of $X$ gates) on the $\rm{B}$ registers and perform the Swap test, without any need for or reference to $\rm{C}$. 
\section{Implementation Circuits} \label{sct:appE}

\subsection*{Spin in a Magnetic Field}
\begin{figure*}[!h]
    \centering
    \includegraphics[width=0.85\textwidth]{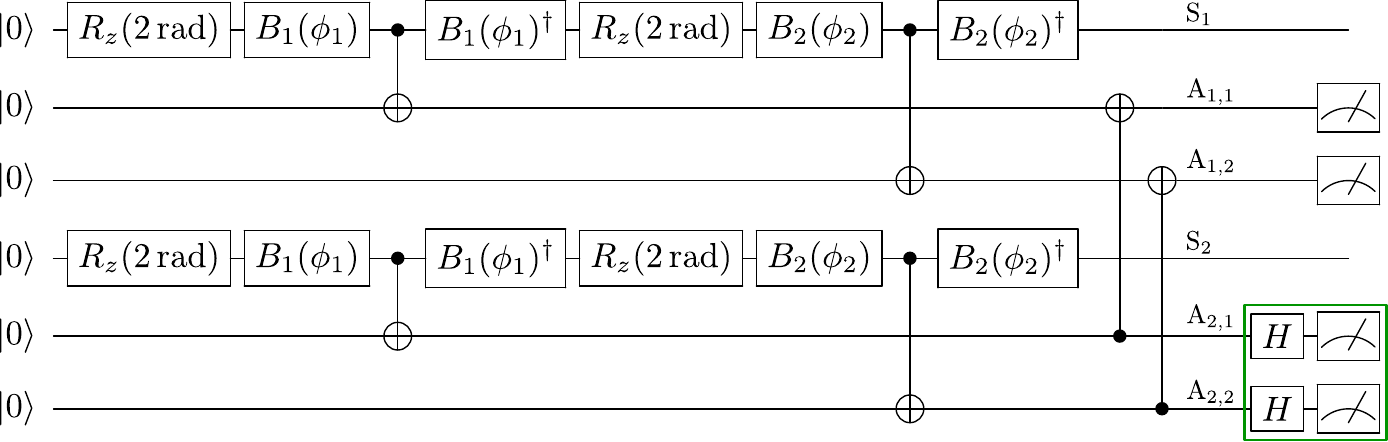}
    \caption{Quantum circuit that we employed to evaluate the cost functions for the spin in a magnetic field. The wires labeled $\rm{S}$ represent the copies of the spin and those labeled $\rm{A}$ represent the ancillas. Note that this circuit prepares two copies of $\sigma^{\rm{A}}$. The gates and measurements inside the solid green box are only included to calculate $\Tr ((\sigma^{\rm{A}})^2)$, as without them this is the circuit to calculate $\Tr (\ZC^{\rm{A}}(\sigma^{\rm{A}})^2)$.}
    \label{fig:B_circ}
\end{figure*}
For our simulations of the spin-1/2 particle in a magnetic field, Fig.~\ref{fig:B_circ} shows the quantum circuit that was used on the simulator and IBM's ibmqx5 processor to perform the cost minimization and to generate the cost landscape plots (shown in Fig.~\ref{fig:B_landscapes}).

\subsection*{Chiral Molecule}

\begin{figure*}[h]
    \centering
    \includegraphics[width=0.9\textwidth]{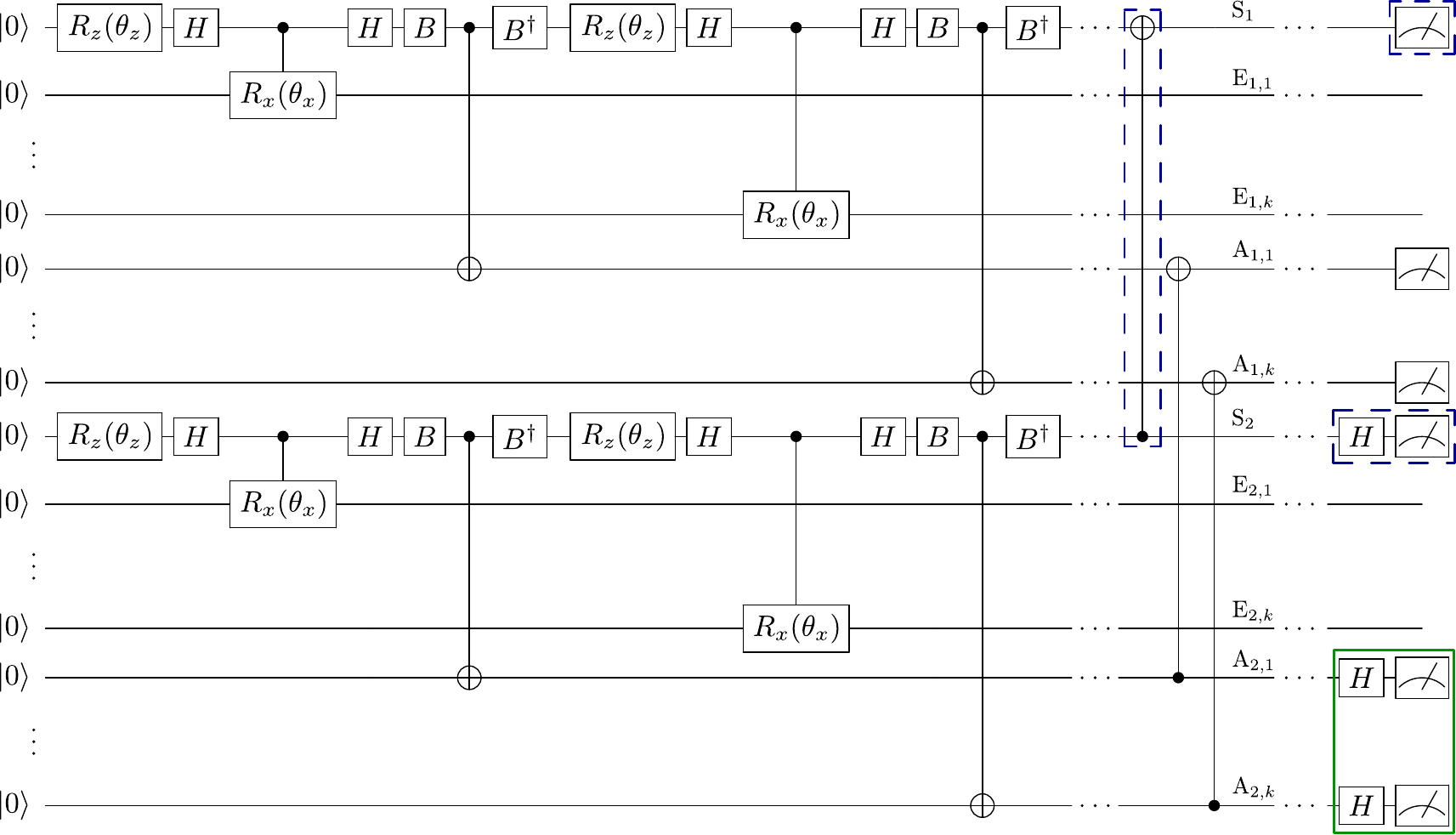}
    \caption{Quantum circuit that we employed to evaluate the cost functions for the chiral molecule example in the main text. The wires labeled $\rm{S}$ represent the chirality degree of freedom of the molecule, $\rm{E}$ represents the environment (other surrounding molecules), and $\rm{A}$ represents the ancillas. Note that this circuit prepares two copies of $\sigma^{\rm{SA}}$ (and hence $\sigma^{\rm{A}}$). The gates and measurements inside the blue dashed boxes are only included when we are evaluating the partial trace cost function (i.e., when working with $\sigma^{\rm{SA}}$ rather than $\sigma^{\rm{A}}$). The gates and measurements inside the solid green box are only included when calculating $\Tr ((\sigma^{\rm{A}})^2)$ or $\Tr ((\sigma^{\rm{SA}})^2)$, and otherwise the circuit calculates $\Tr (\ZC^{\rm{A}}(\sigma^{\rm{A}})^2)$ or $\Tr (\ZC^{\rm{A}}(\sigma^{\rm{SA}})^2)$.}
    \label{fig:chiral_circ}
\end{figure*}

Figure~\ref{fig:chiral_circ} shows the quantum circuit that was used on a simulator to map the cost function landscapes for the chiral molecule (shown in Fig.~\ref{fig:spheres}). The tunneling between the chirality states was modeled as a rotation about the $z$-axis by an angle $\theta_z$. We considered the chiral molecule to be in a gas, and hence its environment is composed of other surrounding molecules that may collide with the molecule of interest. Our model for these collision interactions was implemented by performing a rotation around the $x$-axis by an angle $\theta_x$ (which determines the interaction strength) on an environmental qubit representing the colliding molecule, controlled by the chirality of the molecule of interest. 

\section{Highlighted Applications} \label{sct:appF}
Here we provide a brief outline of two potential applications that would be viable with NISQ computers.

\subsection*{Spin Diffusion}
The phenomenon of spin diffusion has been known for a long time~\cite{original_spin_diff}, but an understanding of the transition from oscillatory dynamics to a classical diffusion equation as system sizes increase is still incomplete. Given that quantum dots with $\sim30,000$ nuclear spins have been shown to exhibit spin diffusion~\cite{Quantum_dot_diffusion}, we expect this to be a very conservative upper bound on the number of spins required.

As a lower bound, simple numerical calculations that we performed show that $\sim 10$ spins do not appear to exhibit spin diffusion. In particular, our calculations showed that, for these small spin systems, the local magnetization does not provide a consistent family of stationary histories. (See the main text for an example of stationary histories for chiral molecules.) Note that the local magnetization forming a consistent family would be a pre-requisite for a random-walk description (and hence diffusive dynamics) of spin magnetization. Combining this lower bound with our upper bound, we expect that the transition is likely to be found with $\sim 10^2$ or $\sim 10^3$ spins.

Applying VCH to this problem would also illuminate the nature and sharpness of the transition. Namely, we anticipate that the transition will involve the disappearance of Rabi oscillations (a signature of quantum interference) for magnetization as the number of spins increases. A natural question is whether such oscillations disappear completely at a critical system size, analogous to how the chirality oscillations disappeared for the chiral molecule (discussed in the main text) at a particular decoherence rate~\cite{Patrick_chiral}. Another possibility is that the transition is gradual, rather than sharp, and that the oscillations are merely suppressed rather than eliminated with system size.

It has been experimentally demonstrated with echo techniques that coherence is maintained during spin diffusion~\cite{diffusion_echo_1,diffusion_echo_2,diffusion_echo_3}. In other words, the classical diffusion equation can be understood to arise from closed-system dynamics rather than open-system dynamics, i.e., as an effect of coarse graining rather than an interaction with the environment. We would therefore only be interested in ansatzes that represent coarse grained spin information on some subset of the spins and neglect environmental effects. 

Given these considerations, we can estimate the number of qubits needed to apply VCH to this situation and look for the sort of random walks that would give rise to diffusion. Let $n_{\text{total}}$ and $n_{\text{voxel}}$ respectively denote the total number of spins and the number of spins in the region we are following the magnetization of (the voxel). Simulating $n_{\text{total}}$ spins requires $n_{\text{total}}$ qubits. In order to implement the projections, we would need to have at most enough qubits to span a space large enough to account for the $n_{\text{voxel}}+1$ possible magnetizations, though this could be coarse grained further. Therefore, to carry out this investigation for $k$ times, we would expect to need roughly 
\begin{equation}
2(n_{\text{total}}+k\lceil\log_2(n_{\text{voxel}}+1)\rceil)
\end{equation}
qubits, where the factor of two comes from the fact we need two copies of the state for VCH. Thus, our estimate for where we expect to find the transition to diffusive behavior with coarse graining translates to needing somewhere around $\sim 10^2$ or $\sim 10^3$ qubits.

\subsection*{Protein Folding}
 Proteins with up to 76 amino acids have been folded thus far using molecular dynamics simulations without adding in external forces to bias the dynamics towards the "correct" configuration~\cite{Atomistic_folding}. However, these simulations do not include decoherence effects and are not capable of fully exploring the vast space of un-biased paths. To move beyond what can be done with these classical tools, we propose to use VCH. 
 
 In order to investigate under which circumstances a protein will follow a single deterministic path or fold by multiple paths, one could implement a quantum simulation of the process using only realistic interaction Hamiltonians and examine the histories. Conjecturing that decoherence by the environment should play an important role, we would need to consider a simulation of an initially unfolded protein as well as its environment. 
 
 Let us consider the simplified case of lattice protein folding for a chain with $n_{\text{AA}}$ amino acids. Each connection between amino acids in such a model can be in $m$ different configurations. This system can be represented with $\lceil(n_{\text{AA}}-1)\log_2(m)\rceil$ qubits. In analogy with the chiral molecule example in the main text, we propose an environment model that would act with different rotations to environment qubits based on the current configuration of each connection, meaning that the size of the environment being modeled would be something like $k\lceil(n_{\text{AA}}-1)\log_2(m)\rceil$ qubits for $k$ times. The size of the ancillas required to record fine grained histories of this system is the same as this environmental size. Finally, given the need for two copies, we end up with a total qubit requirement of
 \begin{equation}
 2(2k+1)\lceil(n_{\text{AA}}-1)\log_2(m)\rceil)    
 \end{equation}
 qubits. For a cubic lattice with $n_{\text{AA}}= 100$ examined at 10 times, this becomes $9,660$ qubits. Given that such a history analysis becomes classically intractable well before the search for the correct (native) configuration does, we therefore think that useful instances of this application will become practical with quantum computers with between $10^3$ and $10^4$ qubits.

\end{document}